# Real time evaluation of overranging in helical computed tomography

Diego Trevisan [a], Faustino Bonutti [b], Daniele Ravanelli [c], Aldo Valentini [a]

[a] Department of Medical Physics, S. Chiara Hospital, APSS Trento, Italy
[b] Department of Medical Physics, S. M. della Misericordia Hospital, Udine, Italy
[c] Department of Proton therapy, S. Chiara Hospital, APSS Trento, Italy

## Abstract

Overranging or overscanning increases the dose delivered to patients undergoing helical Computed Tomography examinations. In order to reduce it, nowadays most of the multidetector tomographs close the X-ray beam aperture at the scan extremes. This technical innovation, usually referred to as dynamic or adaptive collimation, also influences the overranging assessment methods. In particular, the film free approach proposed in previous studies is not suitable for these modern tomographs. The present study aims to introduce a new method of estimating overranging with real time dosimetry, even suitable for tomographs equipped with adaptive collimation. The approach proposed is very easy to implement and time saving because only a pencil chamber is required. It is also equivalent in precision and in accuracy to the film based one, considered an absolute benchmark.

## 1. Introduction

In multislice helical Computed Tomography (CT) X-ray data are interpolated along the cranio-caudal or longitudinal z direction. As a consequence the irradiated range or scan length (L) is larger than the imaged one (PL). This extension of L, usually denoted by overranging (OV), causes a significant additional patient exposure. OV leads to some concern, especially when radiosensitive organs are in close proximity to the scan extremes [1-4]. Moreover, this side effect of multislice helical CT grows with wide beam collimations and high pitch values [5]. For this reason, nowadays almost all modern tomographs reduce OV by means of an adaptive collimation (AC) [6,7]. AC improves the basic safety in CT [8] reducing OV by closing the X-ray beam aperture at the beginning and at the end of the scan, as shown in Fig. 1. OV is always directly measurable by films [9]. However van der Molen [10] and Tien [11] improved OV quantification introducing two film free and time saving approaches. In the "dose-slope" method described in Ref. [10], a dosimeter is placed in the gantry isocentre in a stationary position. Several dose measurements are carried out at different PL values. The dose value at PL = 0 is extrapolated from the experimental data and then converted to OV. Similarly Tien places two dosimeters with high temporal resolution on the couch, exactly at the PL extremes. The time needed by the X-ray beam to cover this distance is measured and then subtracted from the beam-on one. Eventually the result is converted to OV value. Both procedures assume constant beam aperture; for this reason they are not suitable on modern tomographs equipped with AC. The aim of this work is therefore to extend the film free OV evaluation approach to modern AC equipped scanners. In particular this study investigates the use of a standard pencil chamber in OV quantification. To validate the proposed method, all OV estimations provided by pencil chamber are compared with their absolute benchmarks, i.e. OV values measured by films.

The present work is based on Dixon's dosimetric approach [12]: with reference to Fig. 1, the helical CT scan with constant tube current is described by means of the instantaneous dose rate profile $\dot{f}(t, z - vt)$ moving in free air along the cranio-caudal z axis. The parameter v, defined as positive, is the scan speed. The instantaneous dose rate profile $\dot{f}(t, z - vt)$ is function of two variables: the position z and the time t. With reference to Fig. 1, $\dot{f}(t, z - vt)$ explicit dependence on t is due only to the AC opening and closing, and is described by the relation $\frac{\partial \dot{f}}{\partial t} \neq 0$. For the sake of simplicity, Dixon [12] also indicates the midpoint of the scanned range as $z = 0$, and the beam-on time interval in the range $(-t_0/2, +t_0/2)$. With reference to Fig. 1, defining k as the width that instantaneously encompasses the whole $\dot{f}(t, z - vt)$ in the central region of the scan, $\dot{f}(t, z - vt)$ is instantaneously positive in the spatial interval $(vt - k/2, vt + k/2)$. Starting from the translation of the instantaneous dose rate profile, the accumulated total dose profile $D(z)$ [12] at a point z is defined as:

$$D(z) \stackrel{\text{def}}{=} \int_{-t_0/2}^{+t_0/2} \dot{f}(t, z - vt) dt. \tag{1}$$

The accumulated total dose profile $D(z)$ at the midpoint of the scanning range $z = 0$ is defined as maximum dose [12] or central cumulative dose in Ref. [13]. It can be measured as well in dosimetry phantoms as in free air. The central cumulative dose reaches an equilibrium value for a sufficiently long scan length. This equilibrium dose value, denoted by $D_{eq}(0)$ in Ref. [12], is shown in Fig. 2. It is usually measured by placing a short ion chamber at $z = 0$. Dixon [12] also introduces the dose line integral along the z axis as:

$$DLI \stackrel{\text{def}}{=} \int_{-\infty}^{+\infty} D(z) dz, \tag{2}$$

allowing the scan length L to be defined as:

$$L = \frac{DLI}{D_{eq}(0)}. \tag{3}$$

The definition of L above is valid for each kind of scanner, AC equipped or not. As shown in Fig. 2, L is shorter than the total irradiated length i.e. where $D(z) > 0$. The gray area in the same figure refers to DLI.

In this work both DLI and $D_{eq}(0)$ reported in Eq.(3) were measured by means of a pencil chamber (PC) placed at the CT gantry isocentre, in a stationary position. The PC was also uncoupled to the couch displacement. Three different evaluations of L are reported in the following sections. Nominal L ($L_{nom}$) refers to value obtained from the displayed data at CT console, actual L ($L_{act}$) is measured by films, and effective L ($L_{eff}$) is achieved by the proposed method. This new method provides an indirect evaluation of L. $OV_{nom}$, $OV_{act}$ and $OV_{eff}$ are calculated by subtracting the PL from the corresponding L.

## 2. Materials and methods

### 2.1 Tomographs

The $OV_{eff}$ evaluation method was tested on three CT scanners. They were a Philips Brilliance 16S (Eindhoven-NL) installed at the S. Chiara Hospital of Trento, Italy (Tomograph A), a Siemens Somatom Definition AS 128S (Erlangen-DE) operating in the same hospital (Tomograph B), and a General Electric Discovery TM CT750 HD 64S (Fairfield CT-US) installed at the S.M. della Misericordia Hospital, Udine, Italy (Tomograph C). Tomograph A was not equipped with AC. Significant technical differences were also present in AC designs: whereas Tomograph B completely closed the collimators at the scan extremes [6], Tomograph C got only a narrower beam profile [7]. In an attempt to evaluate a wide range of OV values, very different CT scans were considered. As shown in Table 1, a wide range of pitch values and nominal collimations [5] were selected. The nominal dose length product (DLP) and the volume-weighted CT dose index ($CTDI_{vol}$) were also recorded. Its ratio, as indicated in Ref.[14], was considered as the nominal scan length $L_{nom}$. The $OV_{nom}$ was calculated by subtracting PL to L. The PL value, depending on the reconstructed slice thickness, was indicated on the operator's console.

### 2.2 Overranging measurements with films

$OV_{act}$ were previously measured by means of Gafchromic films (GF). GF type XR-QA2 (International Specialty Products, Wayne, NJ, US, lot number A10121202, expiration date 31/10/2014) were used because of their symmetry in sensitivity on 360 degrees [15]. The lot was beforehand calibrated as indicated by Rampado et al. [16]. GF were placed on the table and then irradiated with the scan conditions summarized in Table 1. They were then read by a flat bed scanner (Epson Expression 1000 XL). All scanned GF images were analyzed with an ImageJ (Wayne Rusband, National Institute of Health, US, http://rsb.info.nih.gov//ij version 1.44o) home-made plug-in, providing a D(z) profiles for each GF piece [9]. A $D_{eq}(0)$ measurement was carried out by recording the mean value of each D(z) plateau. The subsequently numerical integration of the same D(z) profile provided the DLI value according to Eq.(2). DLI was then divided by $D_{eq}(0)$ to obtain $L_{act}$, as indicated in Eq.(3). Eventually, $OV_{act}$ value was calculated by subtracting PL from $L_{act}$. This procedure was repeated on five GF pieces for each CT protocol reported in Table 1, allowing the standard deviation of $OV_{act}$ experimental data to be calculated.

## 2.3 Overranging evaluations with pencil chamber

$OV_{eff}$ were evaluated by means of a Radcal 10X5-3CT PC with a charge-collection-length l of 100 mm, coupled to an electrometer (Radcal Corporation 426 West Duarte Road, Monrovia, California 91016, US). Two different PC settings were alternatively available [17]: the PC set in dose rate mode (dr) provided one reading per second, denoted by $\dot{R}_{dr}$ and expressed in mGycm/min units. In dose accumulate mode (da) it provided a reading in mGycm units, indicated as $R_{da}$. In order to collect the whole x-Ray static beam along the cranio-caudal direction, the PC was placed, as shown in Fig. 3, exactly at the gantry isocentre in the stationary longitudinal position. It was also uncoupled to the couch displacement. The PC was then irradiated with the same scan parameters reported in Table 1. Particular care was always taken in avoiding any couch irradiation. Two CT scan repetitions were performed with the PC set in dr mode, and a total of ten $\dot{R}_{dr}$ values were recorded. As explained in Appendix A, all they were only in the central part of the scan, with blocked AC. The PC set in da modality was then exposed three times, always with the same CT scan parameters previously used. A $R_{da}$ measurement was carried out for each scan repetition. As explained in Appendix A, $R_{da}$ and $\dot{R}_{dr}$ were related to DLI and $D_{eq}(0)$ values, respectively. Thus $OV_{eff}$ was quantified, according to Eq.(10), as:

$$OV_{eff} = L_{eff} - PL = F \, v \, \frac{R_{da}}{\dot{R}_{dr}} - PL \,, \qquad (4)$$

with the scan speed v calculated as (nominal collimation x pitch / rot. time). $F = c_{da}/c_{dr}$ is the ratio between the two PC calibration factors. The role of this parameter is addressed in Appendix B. $OV_{eff}$ uncertainty analysis was performed by summing the relative standard deviations of F, $\dot{R}_{dr}$ and $R_{da}$, according to the error propagation theory [18].

## 3. Results

In Table 2 are reported all $R_{da}$ and $\dot{R}_{dr}$ experimental data divided by the pencil chamber length l. These allow evaluating $OV_{eff}$ as indicated in Eq.(4). All $OV_{eff}$ evaluations, together with $OV_{act}$ and $OV_{nom}$ data are reported in Table 3. Both $OV_{eff}$ and $OV_{act}$ values reproduce the $OV_{nom}$ of Tomograph A and B very well, with differences from the nominal values ranging from -1.2 mm up to 1.8 mm, and from -1.0 mm up to 1.4 mm, respectively. $OV_{act}$ and $OV_{eff}$ show on Tomograph C a significant difference from $OV_{nom}$, exceeding 6 mm. It appears that this discrepancy is not related to experimental error but rather to the definition of $L_{nom}$ itself. In particular $L_{nom} \stackrel{\text{def}}{=} DLP/CTDI_{vol}$ has to been revisited as outlined in Ref. [8].

Table 3 and Fig. 4 show also that $OV_{act}$ and $OV_{eff}$ values have a good overlap even for Tomograph C, with absolute discrepancies always less than 2.1 mm. As evident in Fig. 4, the highest OV values of Tomograph A confirm the effectiveness of the AC systems [4,5,19] in dose reduction. This dose reduction is similar to the one related to the tube current modulation [20] and the correct setting of the table high [21]. The lowest OV measured value is also consistent with results obtained in Ref. [5] concerning the OV dependence of scan parameters, i.e. pitch and nominal collimation. $OV_{eff}$ uncertainties reported in Table 3 for Tomograph A without AC, reproduce Tien's results [11], confirming the reliability of the proposed method. Moreover all $OV_{eff}$ evaluations of this study were performed even on modern CT scanners equipped with AC, without any loose in precision or in accuracy. As reported in Table 3, $OV_{eff}$ uncertainties are also similar to $OV_{act}$ ones.

## 4. Conclusions

Films provide a direct measurement of OV allowing the pattern of the accumulated total dose profile D(z) to be known. Nevertheless, several authors proposed alternative real time methods to evaluate OV. All these are cost and time saving, but also not suitable for AC equipped tomographs. The present study overcomes this limitation. In particular it proposes an uniform approach in OV evaluation, i.e. not depending on the presence of AC. The PC based approach described is satisfactory as regards accuracy as well as precision. Moreover it is easy to implement because only a pencil chamber is required. The standard pencil chamber with a charge-collection-length l of 100 mm is not usable for wide collimated scanners with a broad x-Ray collimation extending over the PC charge-collection-length. In such cases the new 300 mm extended pencil chambers, already available on the market, can be used.

## Appendix A: Theoretical background

The instantaneous dose rate profile $\dot{f}(t, z - vt)$ depends on the speed parameter v and is function of the two variables z and t. DLI is evaluable by replacing Eq.(1) into Eq.(2) as $DLI = \int_{-\infty}^{+\infty} \int_{-t_0/2}^{+t_0/2} \dot{f}(t, z - vt) dt\, dz$. By replacing $z' = z - vt$ into the integral above:

$$DLI = \int_{-\infty}^{+\infty} \int_{-\infty}^{+\infty} \dot{f}(t, z') dt dz', \tag{5}$$

where the spatial variable $z'$ is aligned to z axis. $z' = 0$ refers to the gantry isocentre. The right hand side of Eq.(5) does not contain explicitly the translation term vt, i.e. $\dot{f}(t, z')$ describes a static instantaneous dose rate profile. As a consequence DLI can be computed by integrating the $\dot{f}(t, z')$ on its spatial domain $(-k/2, +k/2)$, where k is the width that instantaneously encompasses the whole static $\dot{f}(t, z')$ profile. This interval is always centered at the gantry isocentre as shown in Fig. 3. Thus, a PC with charge collection length $l > k$, i.e. large enough to encompass the whole stationary $\dot{f}(t, z')$ profile, provides a direct measurement of DLI as:

$$DLI = c_{da}\ R_{da}, \tag{6}$$

where $c_{da}$ is the dimensionless calibration factor of the PC operating in da mode, and $R_{da}$ its reading expressed in mGycm, respectively.

As shown in Fig. 2, $D_{eq}(0)$ is the plateau value of $D(z)$ for a sufficiently long scan length. $D_{eq}(0)$ can be obtained by replacing $z = 0$ in Eq.(1) as $D_{eq}(0) = \int_{-t_0/2}^{t_0/2} \dot{f}(t, -vt)dt$. With reference to Fig. 1, the time integration domain of the last integral can be restricted to $(-k/2v, +k/2v)$, i.e. when the instantaneous dose rate profile overlaps the scan centre $z = 0$. Moreover, since the AC is blocked in this central part of the CT scan, then $\frac{\partial \dot{f}}{\partial t} = 0$ for $t \in (-k/2v, +k/2v)$, therefore:

$$D_{eq}(0) = \int_{-k/2v}^{+k/2v} \dot{f}_b(-vt)\, dt, \tag{7}$$

where $\dot{f}_b(z) \stackrel{\text{def}}{=} \dot{f}(t, z)$ for $t \in (-k/2v, +k/2v)$. The suffix b refers to the blocked AC. This propriety is crucial because it allows changing the integration over time in Eq.(7) to a spatial one as:

$$v\, D_{eq}(0) = \int_{-k/2}^{+k/2} \dot{f}_b(z')\, dz', \tag{8}$$

where the spatial variable $z' = -vt$ is aligned to the longitudinal z axis. The right term of Eq.(8) relates to the PC reading in dr mode and suggests that a PC with charge collection $l > k$, positioned as described above for the DLI assessment, allows estimating:

$$v.\, D_{eq}(0) = c_{dr}\ \dot{R}_{dr}. \tag{9}$$

$c_{dr}$ is the dimensionless calibration factor of the PC operating in dr mode and $\dot{R}_{dr}$ the PC reading having units of mGycm/min. The PC reading $\dot{R}_{dr}$ in Eq.(9) refers to the central part of the CT scan at blocked AC. Eventually, by substituting Eq.(6) and Eq.(9) in Eq.(3):

$$L_{eff} = F\, v\ \frac{R_{da}}{\dot{R}_{dr}}, \tag{10}$$

where $L_{eff}$ is the estimation of the scan length L, v is the scan speed, and $F = c_{da}/c_{dr}$ is the ratio between the two PC calibration factors.

## Appendix B: Pencil chamber set up

As explained in Eq. (4) and Eq.(10), $OV_{eff}$ evaluations depend on the ratio $F = c_{da}/c_{dr}$ between the PC calibration factors in da and in dr mode, and not on their absolute values. This section deals with F quantification as ratio between the PC sensitivities in dr and in da mode. With this intention, the PC was positioned 20 mm from a 10 ml glass container with 20 GBq of Technetium-99m. Under these conditions, the PC was irradiated at a high dose rate (about 20 mGy/min) matching $CTDI_{air}$ standard measurements conditions. The PC operating in dr mode provided ten readings in mGycm/min units, indicated as $\dot{r}_{dr}$. After that the PC was set in da mode and a $r_{da}(T)$ measurement (in mGycm units) was carried out for the collection time T. T was chosen equal to 300 s in an attempt to minimize the effect of its uncertainty (±0.5 s). The two measurements described above were repeated ten times, every six minutes. As indicated in Ref.[22], each $r_{da}(T)$ was then converted in $\dot{r}_{da}$ as:

$$\dot{r}_{da} = \frac{\lambda\, r_{da}(T)}{1-e^{-\lambda T}}, \qquad (11)$$

where $\lambda$ is the decay constant of the radionuclide. Note that $\dot{r}_{da} = r_{da}(T)/T$ for $\lambda = 0$.

Since both $\dot{r}_{dr}$ and $\dot{r}_{da}$ were PC readings related to the same source, it was assumed that:

$$c_{dr}\, \dot{r}_{dr} = c_{da}\, \dot{r}_{da}. \qquad (12)$$

Eq. (12) allowed the dimensionless F parameter to be evaluated as:

$$F = \dot{r}_{dr} / \dot{r}_{da}, \qquad (13)$$

i.e. as ratio between the PC sensitivities in dr and in da mode. Fig. 5 shows the F experimental value (1.0013±0.0028), obtained for the PC used in this work.

## 6. Captions

Figure 1 (upper) The adaptive collimation (AC) reduces the X-ray beam aperture at the scan extremes.

(below) The instantaneous dose rate profile $\dot{f}(t, z - vt)$ translates along the cranio-caudal z direction with speed v. It is placed at the scan centre $z = 0$ at $t = 0$, and reaches the position $z = vt$ at $t > 0$. The interval $(vt - k/2, vt + k/2)$ is the spatial range where $\dot{f}(t, z - vt)$ is instantaneously > 0. At the extremes are reported the $\dot{f}(t, z - vt)$ dose rate profiles collimated by the AC. $\dot{f}(t, z - vt)$ change in shape is described by the relation $\frac{\partial \dot{f}}{\partial t} \neq 0$.

Figure 2 The accumulated total dose profile D(z) and its equilibrium value $D_{eq}(0)$ at the scan centre $z = 0$. The gray area represents the dose line integral DLI. The ratio $DLI/D_{eq}(0)$ defines the scan length L. OV is given by subtraction of the imaged length PL from L.

Figure 3 (upper-left) Experimental set up used in real time overranging assessment: the pencil chamber 10X5-3CT with charge-collection-length $l$ of 100 mm was placed exactly at the gantry isocentre ($z' = 0$) in the stationary position. The PC was also uncoupled to the couch displacement.

(right) The static instantaneous dose rate profile $\dot{f}(t, z')$ was completely encompassed in a spatial interval $k < l$ along the $z'$ axis, aligned with the cranio-caudal direction. The pencil chamber in dose rate modality read the highest $\dot{R}_{dr}$ values at blocked adaptive collimation. $R_{da}$ were measured in dose accumulate mode. The gray area indicates $R_{da}$.

Figure 4 Differences between overranging measurements. $OV_{eff}$ refers to pencil chamber measurements, $OV_{act}$ were obtained by gafchromic films.

Figure 5 Ratio between the pencil chamber sensitivity in dose rate and in dose accumulate modalities. The measurements were repeated every six minutes.

## 7. Figures

Fig.1

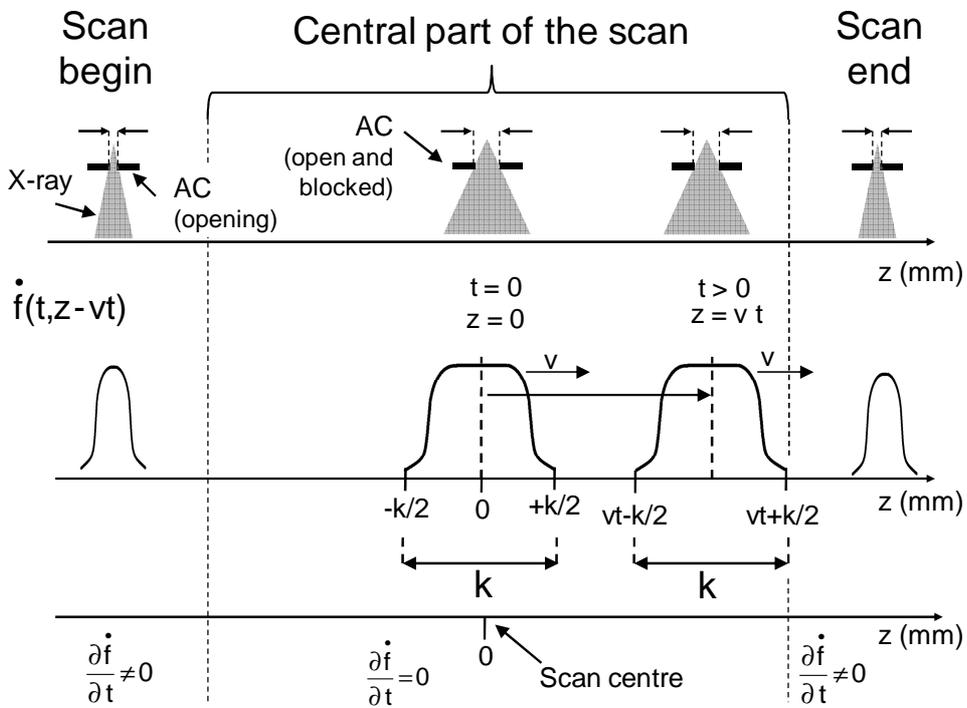

Fig.2

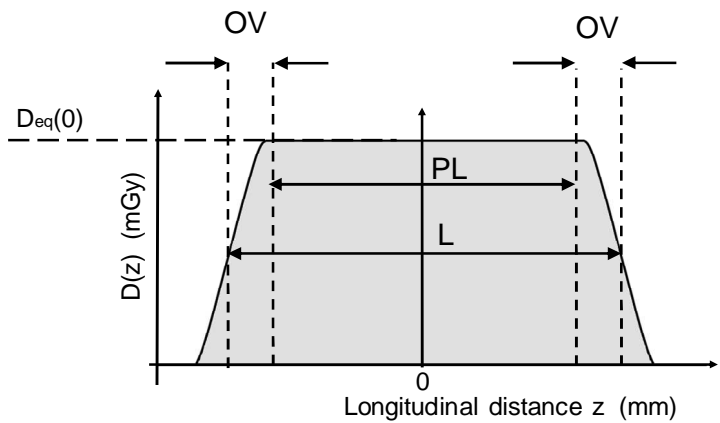

Fig.3

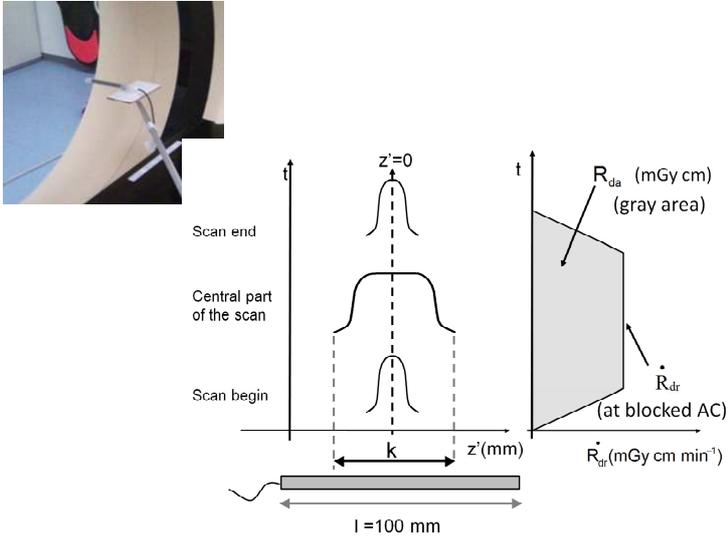

Fig.4

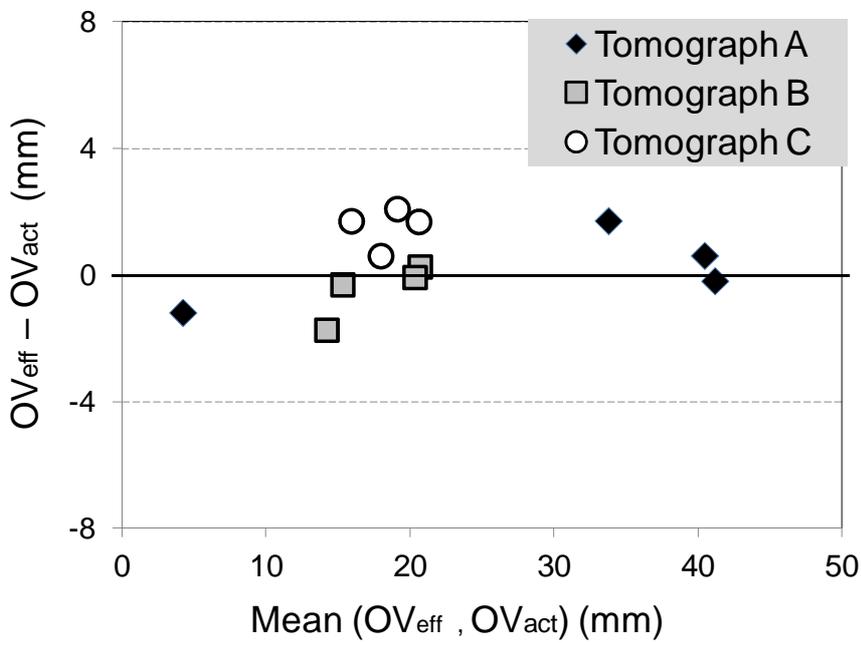

Fig.5

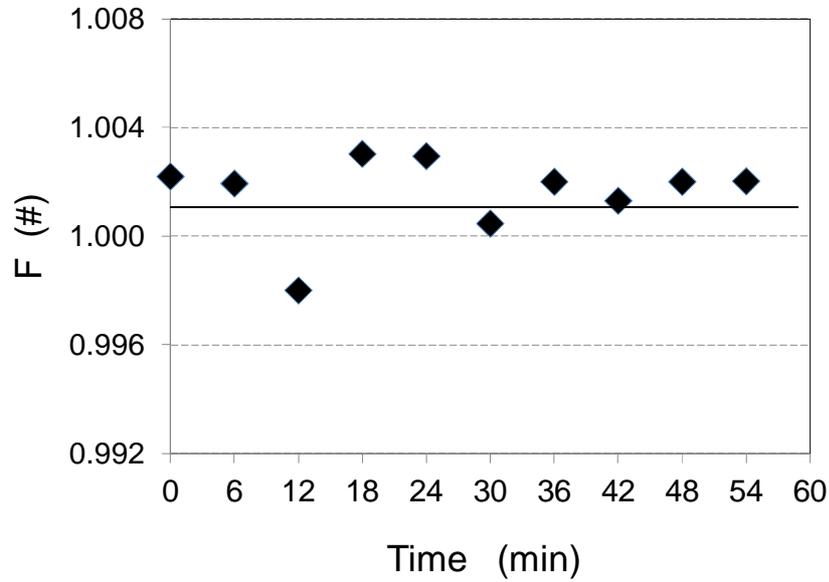

## 8. Tables

Tab.1

**Table 1. Scan parameters (always at 120 kV$_p$) selected for overranging evaluations.**

| Scanner[1] | Protocol | Nominal collimation (mm) | Pitch | Rotation time (s) | Tube current (mA) | Planned Scan length (mm) | Nominal CTDI$_{vol}$ (mGy) | Nominal DLP (mGycm) |
|---|---|---|---|---|---|---|---|---|
| A | Head | 24.0 | 0.688 | 1.5 | 91.73 | 255.0 | 20.3 | 586.7 |
|   | Neck | 24.0 | 0.938 | 1.0 | 187.6 | 261.0 | 10.6 | 318.5 |
|   | Inner ear | 3.0 | 0.35 | 1.5 | 46.7 | 68.0 | 34.5 | 247.5 |
|   | Abdomen | 24.0 | 1.188 | 1.5 | 158.4 | 270.0 | 10.6 | 328.1 |
| B | Head | 38.4 | 0.55 | 1.0 | 165 | 243.1 | 43.15 | 1107.9 |
|   | Neck | 38.4 | 0.80 | 1.0 | 132 | 180.0 | 11.12 | 223.2 |
|   | Inner ear | 38.4 | 0.80 | 1.0 | 144 | 167.0 | 25.70 | 483.9 |
|   | Abdomen | 38.4 | 0.60 | 1.0 | 180 | 232.0 | 19.96 | 492.9 |
| C | Head | 20.0 | 0.531 | 1.0 | 400 | 170.0 | 146.9 | 2809.1 |
|   | Neck | 20.0 | 0.531 | 1.0 | 55 | 170.0 | 20.2 | 396.6 |
|   | Inner ear | 20.0 | 0.531 | 1.0 | 200 | 170.0 | 75.1 | 1437.0 |
|   | Abdomen | 20.0 | 0.969 | 1.0 | 100 | 170.0 | 5.6 | 108.1 |

[1] A: Philips Brilliance 16S; B: Siemens Somatom Definition AS 128S; C: General Electric Discovery TM CT750 HD 64S.

Tab.2

**Table 2. Pencil chamber reads in dose rate and in dose accumulate mode. In brackets are the standard deviations.**

| Scanner[1] | Protocol | $R_{da} / l$ (mGy) | $\dot{R}_{dr} / l$ (mGy/min) |
|---|---|---|---|
| A | Head | 77.92 (0.04) | 177.8 (1.0) |
|   | Neck | 80.92 (0.02) | 362.4 (1.6) |
|   | Inner ear | 33.88 (0.02) | 19.87 (0.12) |
|   | Abdomen | 83.53 (0.04) | 306.4 (1.5) |
| B | Head | 158.3 (0.15) | 783.2 (2.6) |
|   | Neck | 66.08 (0.15) | 607.2 (1.5) |
|   | Inner ear | 66.55 (0.12) | 655.7 (0.9) |
|   | Abdomen | 152.4 (0.07) | 853.4 (2.8) |
| C | Head | 388.8 (0.42) | 1328 (6.1) |
|   | Neck | 55.31 (0.07) | 184.3 (0.9) |
|   | Inner ear | 199.8 (0.14) | 670.3 (3.4) |
|   | Abdomen | 28.35 (0.01) | 175.3 (1.0) |

[1] A: Philips Brilliance 16S; B: Siemens Somatom Definition AS 128S; C: General Electric Discovery TM CT750 HD 64S.

Tab.3

**Table 3. Nominal and measured overranging. In brackets are the standard deviations.**

| Scanner[1] | Protocol | Overranging (mm) | | |
|---|---|---|---|---|
|  |  | Nominal ($OV_{nom}$) | Actual[2] ($OV_{act}$) | Effective[3] ($OV_{eff}$) |
| A | Head | 34.0 | 32.8 (1.1) | 34.8 (1.8) |
|   | Neck | 39.5 | 40.0 (1.3) | 41.0 (1.6) |
|   | Inner ear | 3.7 | 4.8 (0.3) | 3.7 (0.5) |
|   | Abdomen | 39.5 | 41.1 (0.8) | 41.3 (1.8) |
| B | Head | 13.7 | 15.1 (0.6) | 13.4 (1.1) |
|   | Neck | 20.7 | 20.6 (0.3) | 20.9 (0.9) |
|   | Inner ear | 21.3 | 20.4 (0.3) | 20.3 (0.7) |
|   | Abdomen | 14.9 | 15.5 (1.2) | 15.2 (1.1) |
| C | Head | 21.2 | 15.1 (1.2) | 16.8 (1.0) |
|   | Neck | 26.3 | 19.8 (0.4) | 21.5 (1.1) |
|   | Inner ear | 21.3 | 18.1 (0.4) | 20.2 (1.1) |
|   | Abdomen | 23.0 | 17.7 (0.3) | 18.3 (1.2) |

[1] A: Philips Brilliance 16S; B: Siemens Somatom Definition AS 128S; C: General Electric Discovery TM CT750 HD 64S.

[2] Performed with Gafchromic films XR-QA2.

[3] Performed with Pencil-Chamber 10X5-3CT.